\documentclass[12pt,a4paper]{article}%
\usepackage{amsmath}
\usepackage{graphicx}%
\usepackage{amsfonts}%
\usepackage{amssymb}

\begin{document}

\title{New Approach to Sgr A* Problem}
\author{L.V.Verozub\\Kharkov National University}
\maketitle

\begin{abstract}
The hypothesis that radiation of Sgr A* is caused by accretion onto a
supermassive compact object without the events horizon is studied. The main
equations of the accretion and a relativistic equation of the trasfer
radiation are obtained. The synchrotron spectrum in the vicinity of the maximum is considered.

\end{abstract}

\section{Introduction}

An analysis of stars motion in the dynamic center of the Galaxy gives
evidences for the existence of a supermassive ($2.6\cdot10^{6}M_{\odot}$ )
compact object .( See reviews \cite{Goldwurm}, \cite{MeliaFalke01}). There are
three kinds of explanation of the observed perculiarity of the object:

1.The gas accretion onto the central object - a supermassive black hole.

2. The ejection of the magnetized plasma from the vicinity of the Schwarzshild
radius of the above object.

3. The explanations are based on some hypotheses about another nature of the
central objects (neutrino ball, boson stars).

In the present paper we begin an investigation of the assumption that the
radiation of Sgr A* is conditioned by a spherically symmetric accretion onto a
supermassive compact object without the events horizon. The possibility of the
existence of such a kind of objects follows \cite{Verozub95}, \cite{VerKoch00}
from our gravitation equations \cite{Verozub91}$.$ This is a stable
configuration of the Fermi-gas with radius $R$ less than the Schwarzshild
radius $r_{g}$ ( $R$ is about $0.04\ r_{g}$ for the mass $2.6\cdot
10^{6}M_{\odot}$) which unlike a black hole has no the events horizon.

It was shown earlier \cite{VerBan98} that the accretion onto objects of this
kind does not contradict the low observed bolometric luminosity of Sgr A* ($<$
$10^{37}$ $erg/s$ ).

The spectrum of the radiation near its maximum ( $10^{13}\div10^{14}$ $Hz$ ),
that supposedly comes from the vicinity of the object surface, is obtained. It
follows from the results that the analyzed model can be considered as one of
the explanation of Sgr A* radiation.

\section{Accretion equations}

In view of the used bimetric gravitation equations \cite{Verozub91} a remote
observer can consider accretion on the object as a process in Pseudo-Euclidean
space - time.  Euler's equations of the relativistic hydrodynamics in flat
space - time in the presence of a force-field $G^{i}(x)$ are of the form%

\begin{equation}
c^{-2}\nabla_{\alpha}T^{\alpha i}=G^{i},
\end{equation}
where $T^{\alpha i}$ are the components of the energy-momentum tensor
$T^{\alpha\beta}$ of a perfect fluid (Greek indexes run from 0 to 3 and Latin
-- from 1 to 3), $\nabla_{\alpha}$ is the covariant derivative with respect to
coordiates $x^{\alpha}$ in the used coordinate system.

In the case of the spherically symmetric accretion and spherical coordinates
$(t,$ $r,$ $\varphi,$ $\theta)$ we obtain the following differential equation
for the radial component $u$ of the 4-velocity $u^{\alpha}$ of gas
\begin{equation}
wuu^{^{\prime}}+(1-u^{2})P^{^{\prime}}=G^{i}/c^{2},
\end{equation}
where $w$ is the mass-energy density of the gas, $P$ is the pressure, the
prime denotes a derivative with respect to $r.$

The field $G^{i}(x)$ consist of the force density of the radiation pressure
$P_{rad}$ , caused by the accretion, and the density $F_{g}$ of the
gravitational force.

It should be noted that according to \cite{Verozub91} the maximal velocity of
a test particle, falling free onto the object, does not exceed 0.4 of light
velocity. Consequently, the radial component $V$ of the 3-velocity satisfies
the condition $V^{2}/c^{2}\ll1.$

As a result, the Euler equation of the spherically - symmetric accretion onto
the compact object without the events horizon is given by%

\begin{equation}
wVV^{\prime}+c^{2}(P+P_{rad})^{\prime}-F_{g}=0.
\end{equation}
In this equation
\begin{equation}
w=m_{p}c^{2}n+P+\varepsilon,
\end{equation}
where $m_{p}$ is the proton mass, $n$ is the density of particles,
$\varepsilon$ is the internal energy of the gas:%

\begin{equation}
\varepsilon=\alpha nkT+3P_{rad}+B^{2}/8\pi,
\end{equation}
The first righr-hand term is the equation of state of an ionized gas, $k$ is
the Boltzman constant. (In this paper we set the constant $\alpha=3$ at
$10^{5}K<T<6\cdot10^{9}K$ , and $a=9/2$ at $T>6\cdot10^{9}K$ ). The second
term is the density energy of the radiation at the accretion, the third is the
energy density of the magnetic field $B$, frozen in the accretioning plasma.

The density of the gravity force $F_{g}$ can be found from the equation of the
motion of the test particle in a given field. It is is of the form \cite{Verozub91}%

\begin{equation}
F_{g}=c^{2}J_{1}+(J_{2}-2J_{3})V^{2},
\end{equation}
where%

\begin{align}
J_{1}  &  =\frac{C^{\prime}}{2A};\ J_{2}=\frac{A^{\prime}}{2A};\ J_{3}%
=\frac{C^{\prime}}{2C};\\
C  &  =1-\frac{r_{g}}{f};\ \ A=\frac{f^{\prime2}}{C};\ \ f=(r_{g}^{3}%
+r^{3})^{1/3}.
\end{align}

The conservation of mass at the condition $V^{2}/c^{2}\ll1$ yields the equation%

\begin{equation}
4\pi r^{2}\rho V=const,
\end{equation}
where the constant is  the mass accretion rate $\overset{\cdot}{M}$ $=dM/dt$ .

The first equation of thermodynamics yields the equation%

\begin{equation}
d\left(  \frac{w}{n}\right)  =Pd\left(  \frac{1}{n}\right)  +dQ,
\end{equation}
where $dQ$ is an increment of the heat per on a particle, caused by cooling
and heating of the accretioning gas at the interval $dr.$ We take into account
cooling caused by sychrotron and bremsstrahlung radiation, and also by the
converse Compton effect. According to \cite{Pacholchik} we suppose that
contribution to $dQ$ because of the bremsstrahlung (in CGS units) is%

\begin{equation}
dQ_{t}=-1.4\cdot10^{-27}n\sqrt{T}V^{-1}dr,
\end{equation}
and the one because of the sychrotron radiation and reverse Compton effect is
given by%

\begin{equation}
dQ_{s+k}=-(\xi_{1}+\xi_{2})k^{2}T^{2}V^{-1}dr,
\end{equation}
where $\xi_{1}=2.37\cdot10^{-3}B^{2}$ and $\xi_{2}=3.97\cdot10^{-2}\ 3P_{rad}.$

The heating of the accretioning gas is caused by the dissipation of the
magnetic field $B$ as a consequence of maintaining the equality between the
energy of the magnetic field $B$ and the kinetic energy of the matter \cite{Schvartzman}%

\begin{equation}
\frac{B^{2}}{8\pi}=\frac{m_{p}V^{2}}{2}%
\end{equation}
for all $r$, which is generally accepted. The contribution in $dQ$ due to of
it is given by%

\begin{equation}
dQ_{B}=\frac{B^{2}}{8\pi n}\left(  \frac{V^{\prime}}{V}+\frac{2}{r}\right)
dr.
\end{equation}
Finally, we obtain the equation
\begin{equation}
\left(  \frac{w}{n}\right)  ^{\prime}=P\frac{n^{\prime}}{n}+Q_{t}^{\prime
}+Q_{s+k}^{\prime}+Q_{B}.
\end{equation}

\qquad The simultaneous solution of the above equations for a given value of
$\overset{\cdot}{M}$ and magnitudes $n,$ $V$ and $T$ at infinity allows us to
find the functions $n(r),$ $V(r)$ and $T(r)$ , which used for finding the
spectrum of radiation.

\section{Relativistic transfer equation}

The weakest point of the existing models of Sgr A* is not rigorous enough
consideration of the transfer radiation. In the equation an influence of
gravitation on the frequency of the photon and its motion must be taken into
account. Only the relativistic transfer equation can be used for this purpose.
It is a relativistic Boltzman equations for photon gas \cite{Lindkvist},
\cite{Shmidt-Burg}, \cite{Zane}.  We assume that in the spherically symmetric
field the distribution function $\mathcal{F}$ is the function of the radial
distance from the center $r$ , of the frequency $\nu$ and the photon direction
which can be defined by the cosine of the horizontal angle -- $\mu$. (The
spherical coordinate system is used). The relativistic Boltzman equation is
given by%

\begin{equation}
\frac{d\mathcal{F}}{ds}=St(\mathcal{F}),\label{Bolz1}%
\end{equation}
where $d\mathcal{F}/ds$ is the 4-trajectory $x^{\alpha}(s)$ derivative in
space-time with the metric differential form $ds$ and the right-hand member is
a relativistic collisions integral. By using for photons $t$ (or $x^{0}$ ) as
a parameter along 4-trajectory we arrive at the transfer equation in the form
\begin{equation}
\left(  \frac{1}{c}\frac{\partial}{\partial t}+\mathbf{n\nabla+}\frac{d\mu
}{dx^{0}}\frac{\partial}{\partial\mu}+\frac{d\nu}{dx^{0}}\frac{\partial
}{\partial\nu}\right)  \mathcal{F}=St(\mathcal{F}),\label{TransferEq0}%
\end{equation}
where the magnitudes $d\mu/dx^{0}$ and $d\nu/dx^{0}$ must be found from our
equations of the photon motion in the gravitation field \cite{Verozub95} . The
collisions integral is of the form
\begin{equation}
\chi\text{ }(S/\beta-\mathcal{F}),
\end{equation}
where $\chi$ is the absorption coefficient, $\eta$ is the emissivity, $S$
$=\eta/\chi$ is the source function and $\beta$ $=h^{4}\nu^{3}/c^{2}.$ In this
paper we do not take into account light diffusion. An intensity $I$ of the
radiation is related to $\mathcal{F}$ as $I=\beta F.$

\section{Spectrum of the synchrotron radiation of Sgr A*}

We have used the characteristic method \cite{Shmidt-Burg}, \cite{Zane} for the
solution of the transfer equation. Since photon's trajectories are the
characteristics of the partial differential equation (\ref{TransferEq0}), they
are reduced to ordinary differential equations along these trajectories. In
our case the equations are%

\begin{equation}
\frac{d\mathcal{F}}{dr}=\frac{c}{v}\text{ }\chi\text{ }(\frac{S}{\beta
}-\mathcal{F}), \label{TransferEq1}%
\end{equation}
where $v$ is the radial photon velocity.

According to \cite{Verozub91} in the spherically - symmetric field
\begin{equation}
v=c\sqrt{\frac{C}{A}\left(  1-C\frac{b^{2}}{f^{2}}\right)  },
\end{equation}
where $b$ is the impact parameter of photon.

For numerical estimates we set emissivity \cite{Pacholchik}
\begin{equation}
\eta=c_{3}BnF(X),
\end{equation}
where in CGS units $c_{3}=1.87\cdot10^{-23},$ $B$ is the magnetic field, $n$
is the electron density as the function of $r,$
\begin{equation}
F(X)=X\int_{X}^{\infty}K_{5/3}(z)dz,
\end{equation}
$K_{5/3}(z)$ is the modified Bessel function of the second kind, $X=\nu
/\nu_{c},\ \nu_{c}=c_{1}$ $B$ $(kT)^{2}$ , $c_{1}=6.27\cdot10^{18},$ $T$ is
the temperature as a function of $r,$ and $k$ is Boltzman's constant.

The absorption coefficient is approximately
\begin{equation}
\chi=\frac{c_{4}B^{3.2}n}{\nu_{c}^{5/2}}K_{5/3}(X),
\end{equation}
where $c_{4}=4.20\cdot10^{7}.$

The function $n(r)$ is given by
\begin{equation}
n=\frac{\overset{\cdot}{M}}{4\pi m_{p}Vr^{2}}.
\end{equation}

Let $\mathcal{F}_{\nu}(b)$ is the solution $\mathcal{F}_{\nu}(r,b)$ of the
differential equation (\ref{TransferEq1}) for a given $b$ at $r\rightarrow
\infty.$ Then for a distant observer the luminosity at the frequency $\nu$ is
given by%

\begin{equation}
L_{\nu}=8\pi^{2}\int_{0}^{\infty}\beta\mathcal{F}_{\nu}(b)bdb,
\end{equation}

\begin{figure}[ptb]
\begin{center}
\includegraphics[ height=2.5201in, width=3.838in ]{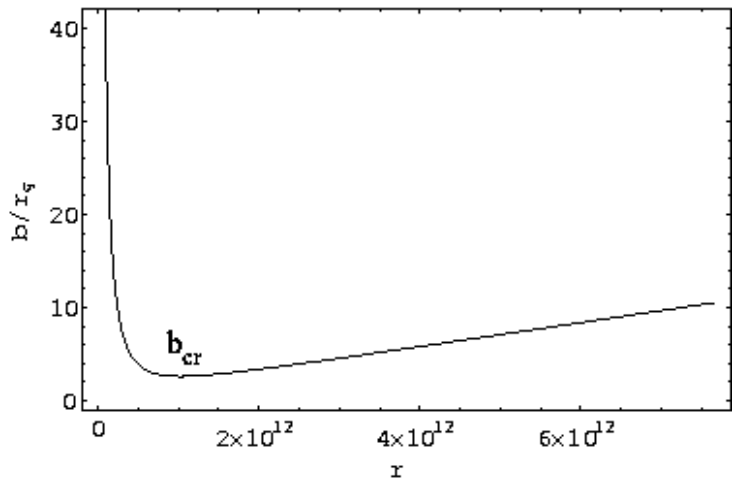}
\end{center}
\caption{The dependence $b$ of $r$ for null-points of the function $v(r)$.}%
\label{bofr}%
\end{figure}

For the correct solution of eq.(\ref{TransferEq1}) it is essential that there
is three types of photons trajectories in the spherically-symmetric
gravitation field in view of the used gravitation equations. It can be seen
from fig. \ref{bofr}. It shows the geometrical locus of the points where the
radial photon velocities are equal to zero which is given by the equation
\begin{equation}
b=\frac{f}{\sqrt{C}} \label{FotonLocus}%
\end{equation}

The minimal value of $b$ is $b_{cr}=3\sqrt{3}r_{g}/2.$ It occurs at the
distance from the center $r_{cr}=\sqrt[3]{19}r_{g}/2.$ The photons whose
impact parameter $b<b_{cr}$ freely extend from the object surface to infinity.
The photons with $b>b_{cr}$ can extend to infinity only if their trajectories
begin at $r>$ $r_{cr}.$ For a given $b$ the corresponding magnitude of $r$ can
be found from eq.(\ref{FotonLocus})

The differential equation for photon trajectories with $b<$ $b_{cr}$ were
integrated at the edge condition $\mathcal{F}(R)=0.$ The distribution function
$\mathcal{F}$ at the points of the curve in fig. \ref{bofr} at $r>r_{cr}$ was
found by solution of differential equation (\ref{TransferEq1}) by used the end
condition $\mathcal{F}=0$ at infinity. Similarly to \cite{Zane} these
magnitudes were used as end conditions for the solution of the
(\ref{TransferEq1}) to find an output radiation.

\begin{figure}[ptb]
\begin{center}
\includegraphics[ height=2.4647in, width=4.0897in ]{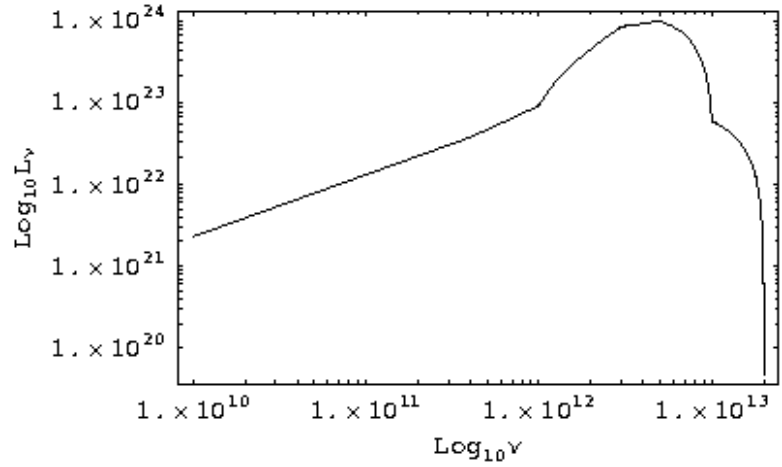}
\end{center}
\caption{The Synchrotron spectrum of Sgr A* near the maximum.}%
\label{spectrum1}%
\end{figure}

Fig.\ref{spectrum1} shows the spectrum of the synchrotron radiation in the
band ($10^{10}\div2\cdot10^{13})$ $Hz.$

\begin{figure}[ptb]
\begin{center}
\includegraphics[ height=2.5477in, width=3.8657in ]{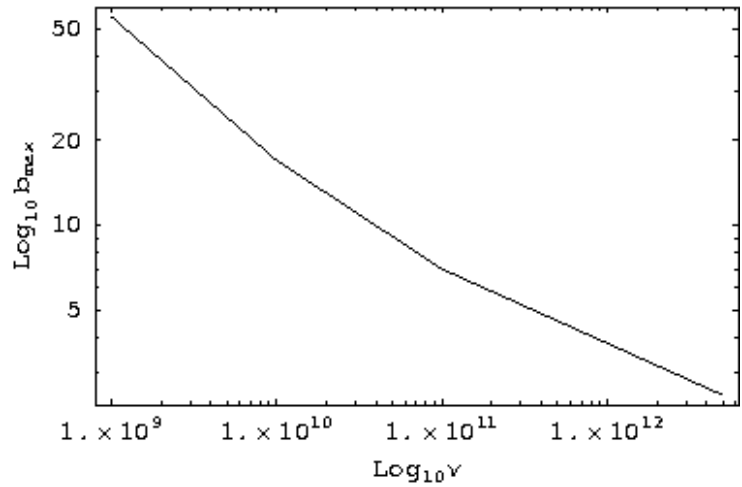}
\end{center}
\caption{The maximal value of the impact parameter vs the frequency.}%
\label{bmax}%
\end{figure}

In the studied frequencies band the functions $\mathcal{F}_{\infty}(b)$ are
weakly depending on $b$ up to some maximal magnitude $b_{\max}$, and after
that quickly decreases. In fig. \ref{bmax} the dependence of $b_{\max}$ on the
frequency is shown.

A comparison (\ref{spectrum1}) and (\ref{bmax}) with the observation data
\cite{Goldwurm}, \cite{MeliaFalke01} shows that the results do not contradict
observations and, therefore, can be a basis for a deeper investigation.

\section{Conclusion}

To clear up question about the nature of Sgr A* is issue of the day since the
problem is the nature of the supermassive compact objects. Since 1992 a number
of models have been created that are able to explain Sgr A* spectrum. It is
the possibility of an ambiguous explanation of the spectrum that means that we
need for more rigorous methods of the calculation of the Sgr A* spectrum. It
is possible only by a simultaneous solution of the relativistic hydrodynamics
and transfer equations.

I am very grateful to S.Zane for a kind explanation of their approach to the
transfer equation and J. Schmid-Burg for sending me his paper on this subject..

\end{document}